\documentstyle[aps,prl,multicol,graphicx]{revtex}

\begin{document}
\input{epsf}

\title{
Electronic properties of antidot lattices fabricated by atomic force
lithography
}
\author{A. Dorn, A. Fuhrer, T. Ihn, T. Heinzel, and K. Ensslin}
\address{Solid State Physics Laboratory, ETH Z\"{u}rich, 8093
Z\"{u}rich,  Switzerland\\
}
\author{W. Wegscheider}

\address{Angewandte und Experimentelle Physik, Universit\"at
Regensburg, 93040 Regensburg, Germany\\
}
\author{M. Bichler}
\address{Walter Schottky Institut, TU M\"unchen, 85748 Garching,
Germany\\
}
\date{\today}
\maketitle
\begin{abstract}

Antidot lattices  were fabricated by atomic force lithography using
local oxidation.
High quality finite 20 x20 lattices are demonstrated with periods of
300 nm.
The low-temperature magnetoresistance shows well developed
commensurability oscillations as well as a quenching of the Hall
effect
around zero magnetic field.
In addition, we find B-periodic oscillations superimposed on the
classical commensurability
peaks at temperatures as high as
1.7 K.
These observations indicate the high electronic quality of our
samples.
\end{abstract}


\begin{multicols} {2}
\narrowtext
Experiments on lateral superlattices have revealed a variety of
unexpected phenomena in the classical as well as in the quantum
regime. Antidot lattices are particularly appealing  for studying
how a classically chaotic system develops quantum
signatures for shorter lattice periods
\cite{Ensslin90,Weiss91,Lorke91,Weiss93,Schuster94,Schuster94b}.
Most high-quality samples have so far been fabricated with electron
beam
lithography. This technology is well-developed and offers reliable
performance as well as excellent electronic quality of the fabricated
samples
down to lattice periods of about 100 nm \cite{Schlosser96,Albrecht01}.
The quest for the experimental observation of a quantum mechanical
bandstructure
based on the periodic lateral potential modulation is ongoing and
requires lattice periods of the order of the Fermi wavelength of the
electrons in the two-dimensional electron gas. Lattice periods of the
order of 40-50 nm are difficult to fabricate by electron beam
lithography  due to the proximity effect.

Lithography based on scanning probe techniques has
evolved into a powerful tool  for defining ultrasmall patterns on
surfaces. \cite{Marrian93} Electronically functional lateral
superlattices have
been fabricated by using the (sharpened) tip of an atomic force
microscope to either scratch a pattern in a resist layer
\cite{Wendel95}
or directly
scratch into the surface of an InAs heterostructure \cite{Cortes98}. An
alternative approach relies on the local oxidation of a water covered
surface by applying a voltage between a conductive AFM tip and the
underlying surface. \cite{Becker87} If applied to AlGaAs
heterostructures a
two-dimensional electron gas close enough to the sample surface (less
than 50 nm for practical considerations) has been shown to be depleted
below the oxidized lines. \cite{Held98} This procedure has the
desirable property that the pattern transfer occurs simultaneously
with the lithography step and no further treatment/processing is
needed. It
 has been  demonstrated that  electronically tunable quantum wires
\cite{Held99} and quantum dots
\cite{Luscher99} can be fabricated with this technology. Important
features of nanostructures fabricated this
way are the smooth \cite{Heinzel00} as well as steep \cite{Fuhrer01}
potential
walls. These are crucial ingredients for the successful
realization of lateral superlattices.


We started from a high quality two-dimensional electron gas with a 4K
mobility of
$\mu = 800'000 cm^{2}/Vs$ and an electron density of
$N_{s} = 5 \cdot 10^{11} cm^{-2}$. The electron gas embedded in the
AlGaAs
heterostructure was 34 nm below the sample surface. For lithography
we used a
commercial Digital Instruments Dimension 3100 with metrology head. 
This AFM  is equipped
with capacitive sensors
 that reduce long time drifts and creep through a hardware feedback and enable
a lateral positioning of the tip with a precision in the nm range.
The humidity of the
atmosphere surrounding tip and sample is kept at a value of about
$40\%$
during the writing process. Typical voltages for the oxidation of the
GaAs
surface were   -10 to -30 V applied to the tip with
the GaAs grounded. The oxide pillars in this structure were
written in tapping mode with a stationary 3s/-18V puls at each site.
The pillars show a high uniformity and have a cone like shape with a
base of about 250nm and a height of about 8nm.

For
large antidot arrays we observed a weakening of the oxidation process
either caused by a deterioration of the AFM tip and/or an insufficient
supply of water on the surface as many oxide pillars are formed.
In this paper we present results on a 20x20 array where we are safely
within the limit of a uniform pattern. 

After the AFM process the sample is covered with a 
uniform metallic layer which allows us to tune the Fermi energy 
of the lattice and in addition protects the sample surface from
contamination or further oxidation.
It is well known that the piezos which drive the scanning unit have
intrinsic creep and hysteresis properties. While the results of these
unwanted features can be eliminated to some degree by software in the
imaging mode, they can have detrimental consequences for lithography.
As long as single nanostructures are envisioned, such as dots and
wires,
small imperfections of their shape can be tolerated since the
potential
landscape has some random background due to the modulation doping
anyway.
For periodic lattices, however, creep and hysteresis can lead to
errors of
the order of the lattice period for arrays with periods below 100 nm.
It is therefore crucial for the precision of our pattern
that we have hardware control
over the actual position of the AFM tip as described above.

Another limitation for the precision of periodic lattices fabricated
by AFM
lithography is thermal drift which cannot easily be compensated by
hardware
since it affects the entire instrument.

\begin{figure}
\caption{(a) Overview of the  entire AFM patterned area. The bright
areas mark
the oxidized regions below which the electron gas is depleted.
The contacts are numbered 1-4.
(b) Close-up of a  3X4 period segment from (a).
(c) Line scan along the direction indicated in (b).}
\end{figure}

We found empirically that a carefully
controlled lab environment ($\Delta T  < 0.5 K$) in combination
with settling times of the
order
of several hours reduce the thermal drift to values which are no
longer
detectable in our antidot lattices.

Figure 1 (a) shows an overview of the entire 20x20 array with a
lattice period of 300 nm.
The bright areas are patterned with the AFM and define the regions of
depleted electron gas. The four outer bars are necessary to define a
measurement
geometry for the finite lattice. Typical breakdown voltages between
areas
of 2DEG separated by insulating lines are $\ge 1 V$ at $T=4.2 K$.
The close-up presented in Fig. 1 (b) as well as the line scan
in (c) show the precision
obtained with this technology. From an analysis of the topography of
the
entire lattice we estimate
the error in antidot position to be $\Delta a/a \approx 1\%$, and
the variation in antidot size to be less than $5\%$.


\begin{figure}
\caption{ 4-terminal magnetoresistance (left hand scale) and
 Hall resistance (right hand scale) at T=1.7 K
 through the finite antidot
geometry, contacts are numbered as
 indicated in Fig. 1 (a) . The inset shows a close-up of
the Hall effect
 around B=0.}
\end{figure}

Figure 2 shows the magnetoresistance of the structure
presented in
Fig. 1 at T=1.7 K.
The resistance $R_{11,44}$ is a 2-terminal resistance
with respect to the square geometry. However, there are
two independent Ohmic contacts in each of the corresponding
electron gas regions next to the square geometry in order 
to eliminate contact resistances. 
The magnetoresistance measured this way displays the well known
commensurability
maxima which are related to classical cyclotron orbits around 1
($B \approx 0.8 T$) resp. 4 ($B \approx 0.2T$)  antidots.
At high magnetic fields where the cyclotron diameter becomes much
smaller than
any of the lattice features, Shubnikov-de Haas oscillations are
recovered.
The Hall resistance $R_{14,23}$ displays plateau-like features close
to the commensurability conditions and a quenching around B=0. All of
these
features are well known from previous experiments.
\cite{Weiss91,Schuster94b}
The quality of the features as presented in Fig. 2 shows that the
electronic
potential landscape created by AFM lithography is at least comparable
to those obtained by conventional electron beam lithography. From the
shape
of the classical maxima in the magnetoresistance as well as from the
onset of Shubnikov-de Haas oscillations we estimate the electronic
size of an antidot at the Fermi energy to be less than 200 nm.
This is consistent with the lithographic size as well as with the
lateral depletion
length obtained for wires \cite{Held99} and dots \cite{Fuhrer01}
fabricated
with the same technology.


\begin{figure}
\caption{Three curves show the resistance at
T=4.2 K (top curve, dashed line), at T=1.7 K (solid line,
vertically offset by 0.5 $k\Omega$), and the difference (bottom curve).
Thin vertical lines connect minima in the magnetoresistance with the very
top curve
where the minima are plotted according to their number which is
extrapolated from filling factors extracted from
the high-field Shubnikov-de Haas oscillations. The
straight dashed line  shows the behavior expected for
consecutively adding
flux quanta through a circle with diameter 300 nm.}
\end{figure}

In addition to the classical effects we observe fluctuations and
oscillations
of phase coherent origin superimposed on the commensurability
maximum. In order to show these effects most clearly the carrier density
is reduced by applying a negative gate voltage
of
$V_{g}=-100 mV$.
Figure 3 shows $R_{11,44}$, 
where the point contacts still support several 1D
modes,
but the antidots have become large enough at the Fermi surface, that
the magnetoresistance maximum corresponding to a ballistic orbit
around 4 antidots is eliminated.
In this regime it  has been demonstrated that B-periodic
oscillations can be observed in an infinite lattice \cite{Weiss93}.
For a finite lattice much smaller than ours, on the other hand, these
quantum oscillations were dominated by ballistic conductance
fluctuations \cite{Schuster94}. In our case where the phase coherence
length is much larger than the lattice period but probably comparable
or smaller than the extent of the
finite lattice, we observe at T=1.7 K both types
of phase coherent phenomena. In
Fig. 3 we show $R_{11,44}$ at
T=4.2 K (dashed line), at T=1.7 K (solid line,
vertically offset by 0.5 $k\Omega$) and the difference between these
two curves (solid curve). Coming from high magnetic fields
Shubnikov-de Haas oscillations dominate the magnetoresistance. By numbering
the minima with appropriate filling factors (the 2D density is
$N_{s} = 3.1 \cdot 10^{11} cm^{-2}$) we continue this procedure
down to magnetic fields
$B \le 1 T$ where the magnetoresistance minima no longer follow a
$1/B$-periodicity. The corresponding positions are plotted in the upper
part of Fig. 3.
In the regime where the electrons classically encircle  a single antidot,
the orbits can no longer contract for increasing magnetic field, since
the Lorentz force is increasingly compensated by electrostatic
deflection at the antidot pillars. In this regime the magnetoresistance
minima follow a $B$- rather than a $1/B$-periodicity.
\cite{Weiss93,Richter95,Hackenbroich95}
Because our lattice is finite, additional features occur in the
magnetoresistance due to ballistic phase coherent
fluctuations \cite{Schuster94}.


We have demonstrated that high-quality antidot lattices
can be fabricated by AFM-mediated local oxidation of
AlGaAs heterostructures. The experimentally observed features,
e.g. classical commensurability oscillations superimposed by phase
coherent fluctuations, demonstrate that the electronic
potential landscape can at least compete with the best samples
prepared by other means. Since our technological approach is
based on scanning probe techniques it offers the intrinsic advantage
to fabricate samples with substantially smaller lattice periods.
This requires high-quality 2DEGs close to the sample surface.
Furthermore, one of the major challenges is to keep the uniformity
of the pattern for smaller periods and thinner oxide pillars.

We are grateful to the Swiss Science Foundation
(Schweizerischer Nationalfonds) for financial support.

\end{multicols}

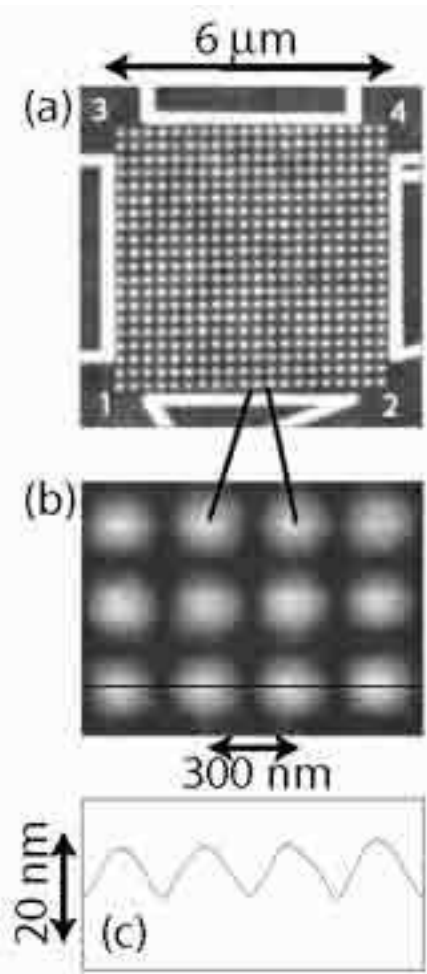
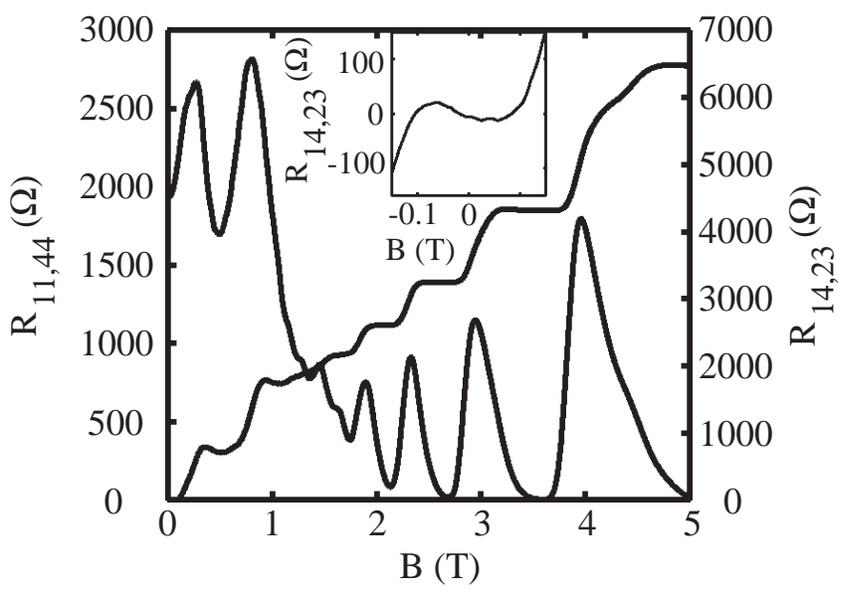
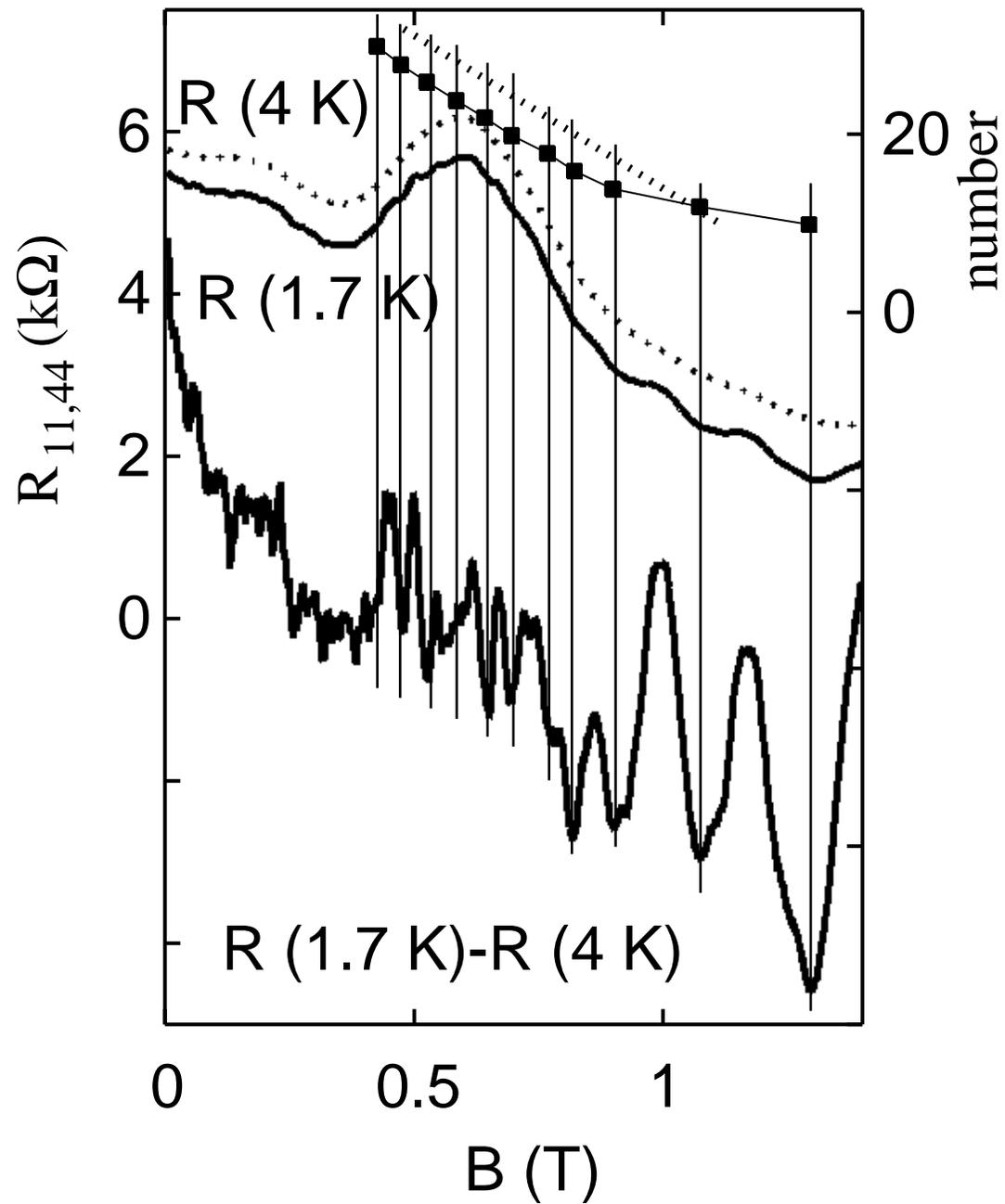
\end{document}